\documentclass[12pt]{iopart}
\usepackage{iopams}  
\usepackage{graphicx}

\begin{document}

\title[Hybrid Josephson junctions on Ba-122 single crystals]{Preparation of hybrid Josephson junctions on Co-doped Ba-122 single crystals}

\author{D~Reifert$^1$\footnote{Present address: Physikalisch-Technische Bundesanstalt (PTB), Bundesallee 100, 38116 Braunschweig, Germany}, N~Hasan$^1$, S~D\"oring$^1$, S~Schmidt$^1$, M~Monecke$^1$\footnote{Present address: Chemnitz University of Technology, Physics Department / Semiconductor Physics, Reichenhainer Strasse 70, 09126 Chemnitz, Germany}, M~Feltz$^1$, F~Schmidl$^1$, V~Tympel$^1$, W~Wisniewski$^2$, I~M\"onch$^3$, T~Wolf$^4$, P~Seidel$^1$}

\address{$^1$Institute of Solid State Physics, Friedrich-Schiller-University Jena, Helmholtzweg 5, 07743 Jena, Germany\\
$^2$Otto-Schott-Institute of Materials Research, Friedrich-Schiller-University Jena, Fraunhoferstrasse. 6, 07743 Jena, Germany\\
$^3$Institute for Integrative Nanosciences, IFW Dresden, Helmholtzstrasse 20, 01171 Dresden, Germany\\
$^4$Institute of Solid State Physics, Karlsruhe Institute of Technology, 76021 Karlsruhe, Germany}
\ead{sebastian.doering.1@uni-jena.de}

\begin{abstract}
In this paper we present a method for processing a hybrid Josephson junction on Co-doped BaFe$_{2}$As$_{2}$(Ba-122) single crystals with a thin film Pb-counter electrode and a barrier layer of TiO$_x$. This includes the leveling and polishing of the crystals and structuring them with thin film techniques such as photo lithography, sputtering and ion beam etching (IBE). The junctions show hysteretical resistively and capacitively shunted junction (RCSJ)-like $I$-$V$ characteristics with an $I_{\mathrm{c}}R_{\mathrm{n}}$-product of about 800\,$\mathrm{\mu}$V.

\end{abstract}

\maketitle
\twocolumn
\section{Introduction}
Single crystals are an ideal tool to investigate the basic properties of a new material due to their nearly perfect crystalline structure. On the other hand Josephson junctions are a good method to investigate the electrical properties of a superconductor. Despite this, only little literature about Josephson junctions on pnictide single crystals\cite{Zhang2009a,Zhou2008,Zhang2009} could be found by the authors. This could be due to the surface properties of pnictide single crystals, e.g. roughness and degradation in a normal atmosphere\cite{Plecenik2013}, and the difficult in-situ preparation of artificial barriers. In the literature mentioned above the single crystals used as one electrode were coupled by point contacts, thick films or even crossed by other single crystals, making the junction area and barrier type difficult to control. Our approach, in principle, enables us to produce controllable junctions with areas ranging from 5$\times$5\,$\mathrm{\mu}$m$^2$ to 50$\times$50\,$\mathrm{\mu}$m$^2$ using normal conducting and insulating barriers, which is advantageous for future systematic investigations of the junction properties. We typically  used a junction area of 15$\times$15\,$\mathrm{\mu}$m$^2$ for the investigation of TiO$_x$ as an insulating material. Polishing the crystal surface to a RMS roughness $\leq5\,$nm is necessary in order to prepare these junctions using photo lithography, sputtering and IBE.
While there are some publications concerning the polishing of different unconventional superconductor materials \cite{Shapoval2008,Strand2009,Michalowski2012}, it is necessary to consider special requirements for the polishing procedure for the preparation of devices on pnictides, which were not fully met in the mentioned works. Ba-122 is sensitive to water and acetone meaning the polishing solvant and the cleaning steps had to be adapted.

\section{Sample Preparation}
\begin{figure}[htb]
\centering
\includegraphics[width=1\linewidth]{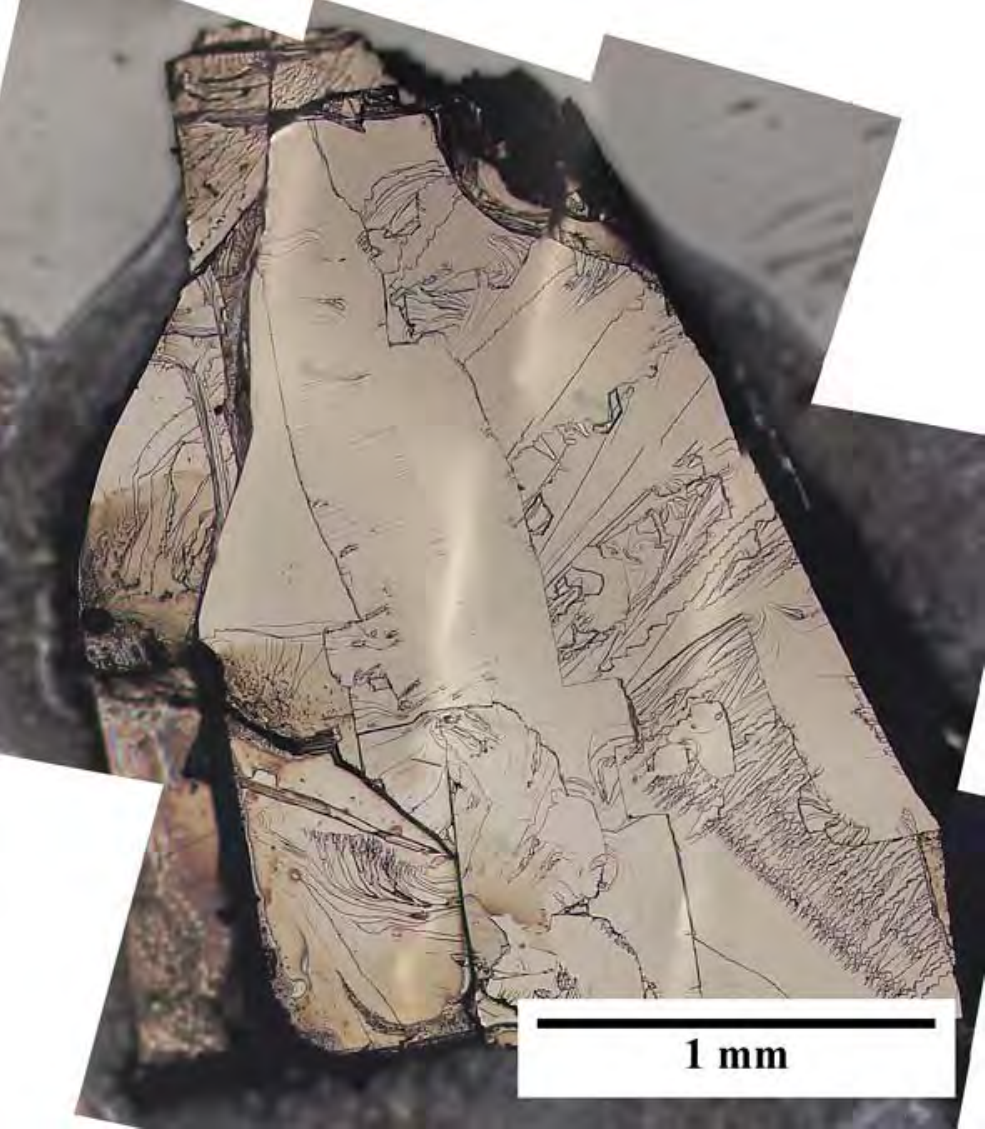}
\caption{Microscope image of a Ba-122 single crystal as-grown.}
\label{fig:sckit06_roh_10x}
\end{figure}

\begin{figure}[htb]
\centering
\includegraphics[width=1\linewidth]{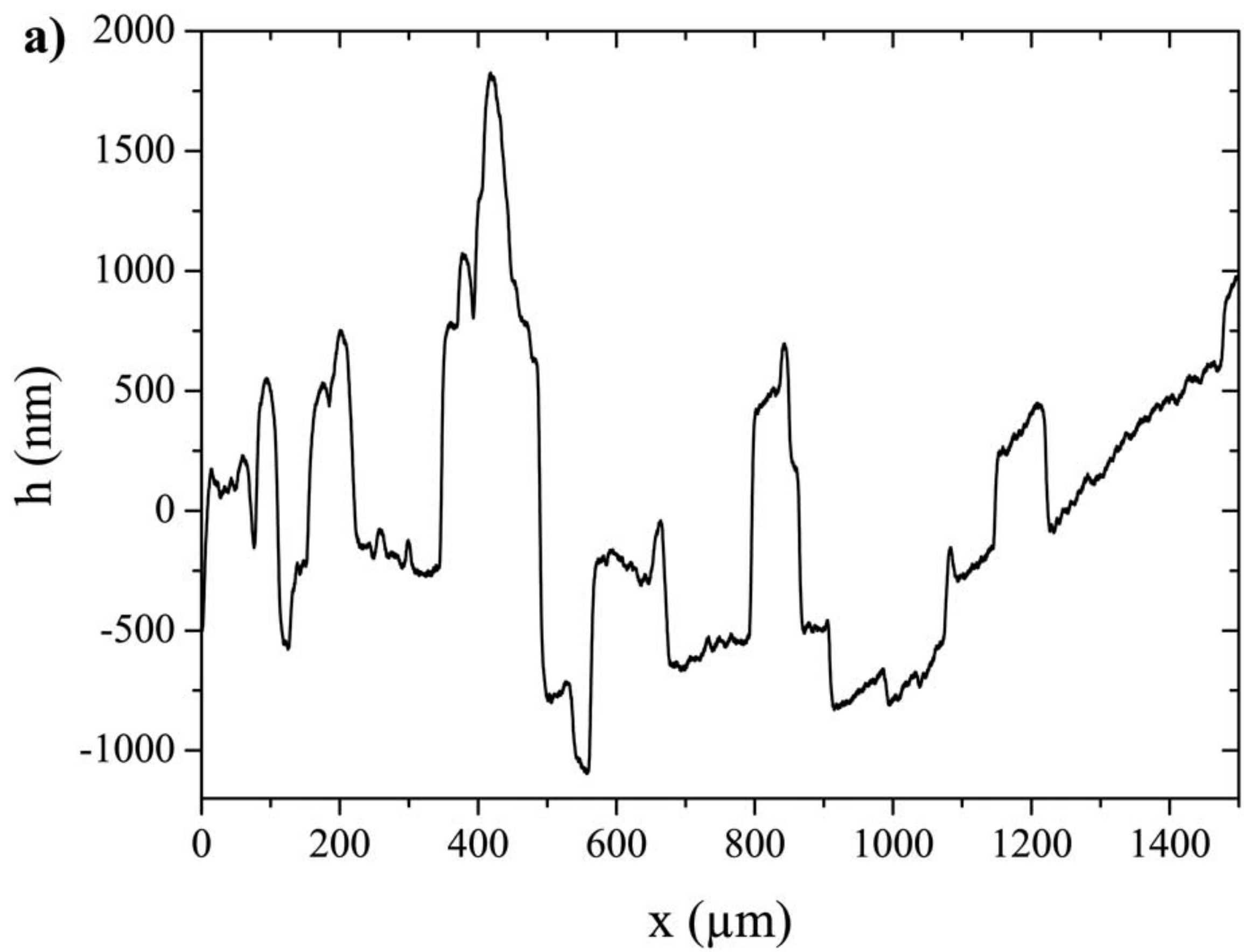}
\includegraphics[width=1\linewidth]{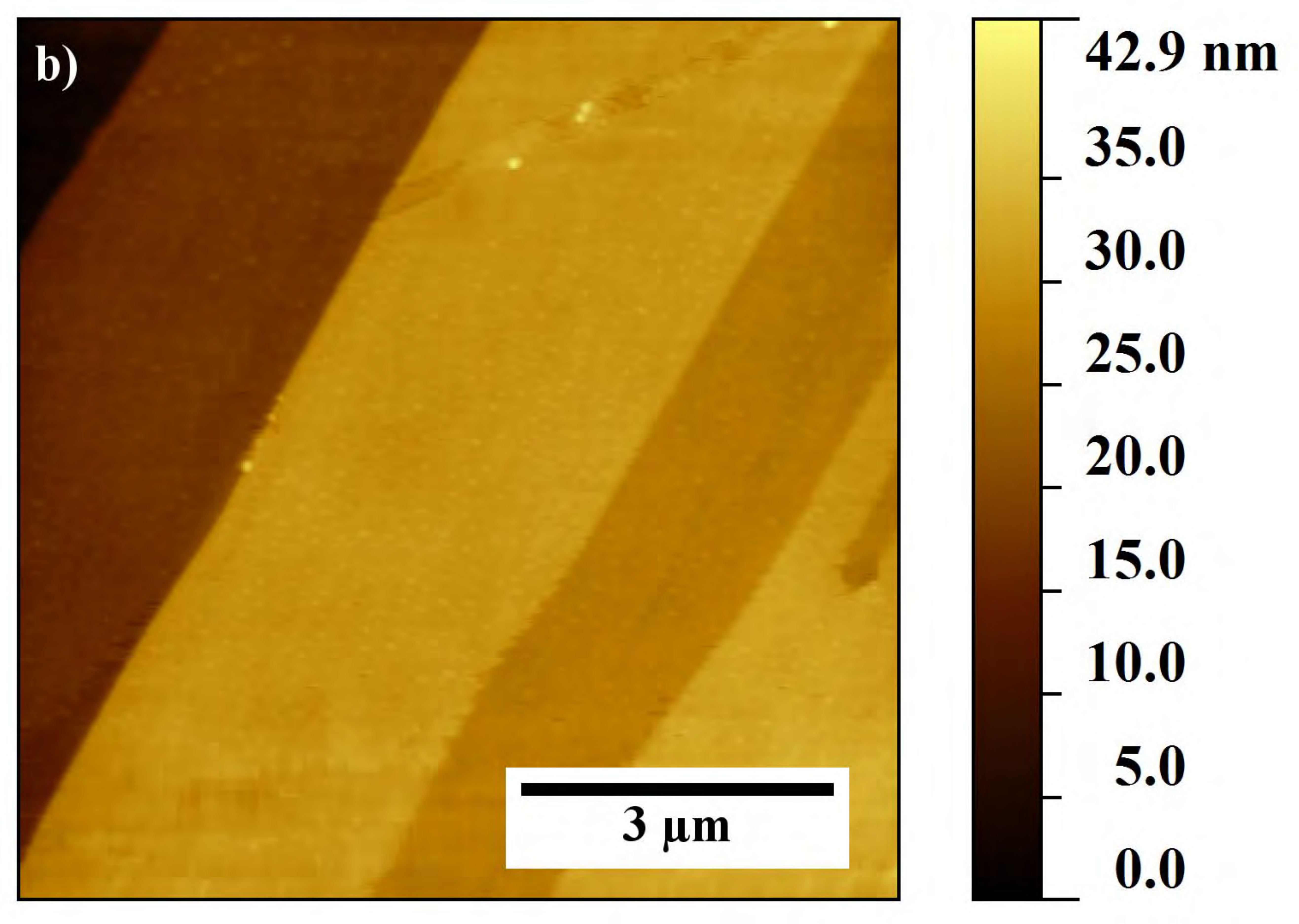}
\caption{a) Topographical profile of the unprocessed surface with a step height of up to several $\mathrm{\mu}$m. b) AFM measurement of the same surface featuring additional nanoscopic steps.}
\label{fig:roh03}
\end{figure}

\begin{figure}[htb]
\centering
\includegraphics[width=1\linewidth]{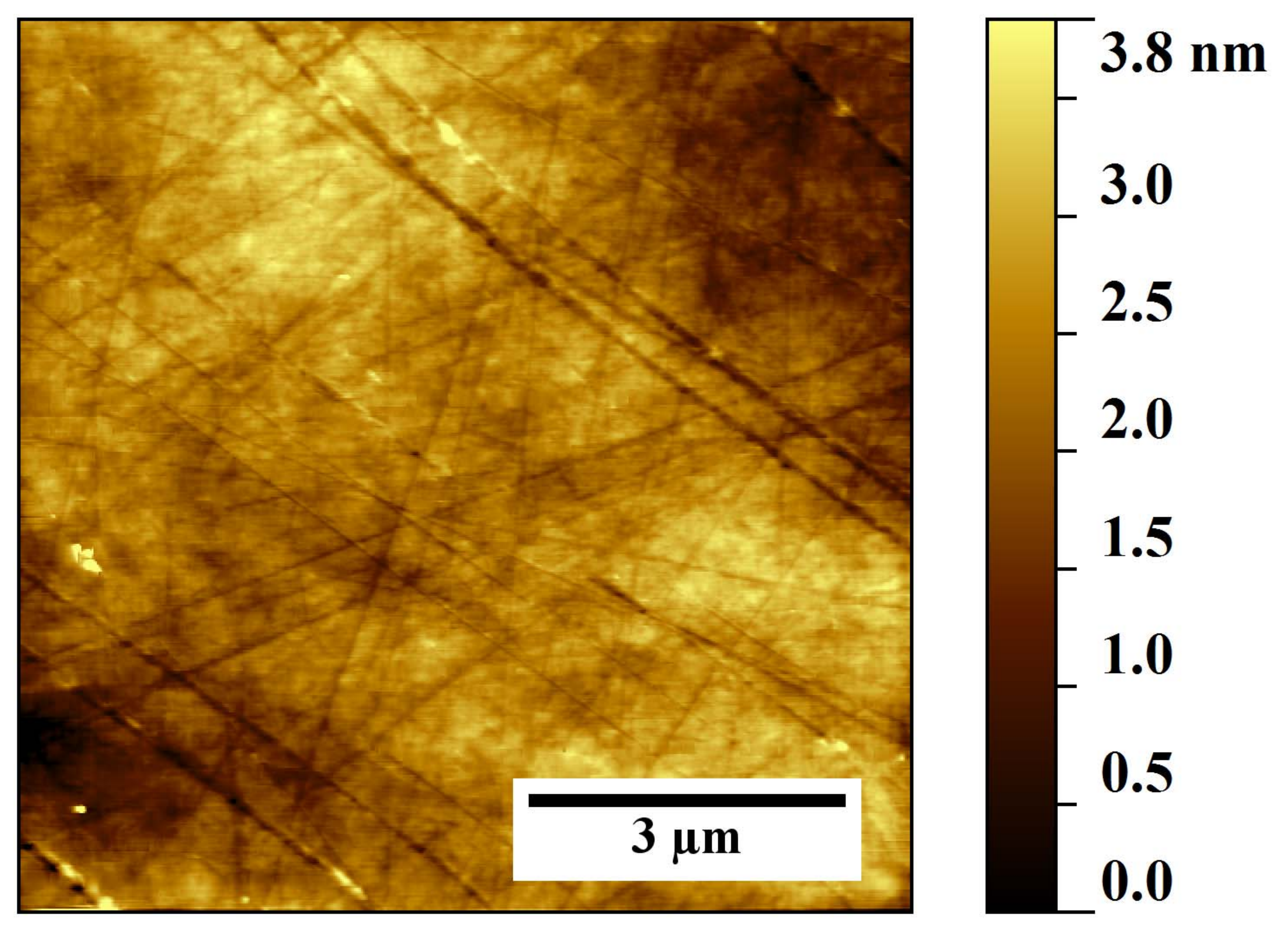}
\caption{AFM measurement of a polished crystal surface with a RMS roughness of 0.6\,nm.}
\label{fig:pol}
\end{figure}

The Ba(Fe$_{1-x}$Co$_{x}$)$_2$As$_2$ crystals we used were grown from self-flux in glassy carbon crucibles in analogy to references \cite{Hardy2009,Hardy2010}. A particularly low cooling rate of 0.3\,$^{\circ}$C/h was applied to minimize the amount of flux inclusions and crystal defects. The composition of the crystals was determined by energy dispersive X-ray spectroscopy to be Ba(Fe$_{1-x}$Co$_{x}$)$_2$As$_2$ with $\mathrm{x}$\,=\,0.054. A critical temperature $T_{\mathrm{c}}$ of 23.5\,K was inductively measured.
Figure \ref{fig:sckit06_roh_10x} shows an optical micrograph of a typical unprocessed single crystal. The surface shows significant structures covering the main plateau. The measurements with atomic force microscopy (AFM) and a stylus profilometer, presented in \fref{fig:roh03} show, that these are steps with a height of several nm to several $\mathrm{\mu}$m. 
\begin{figure}[htb]
\centering
\includegraphics[width=1\linewidth]{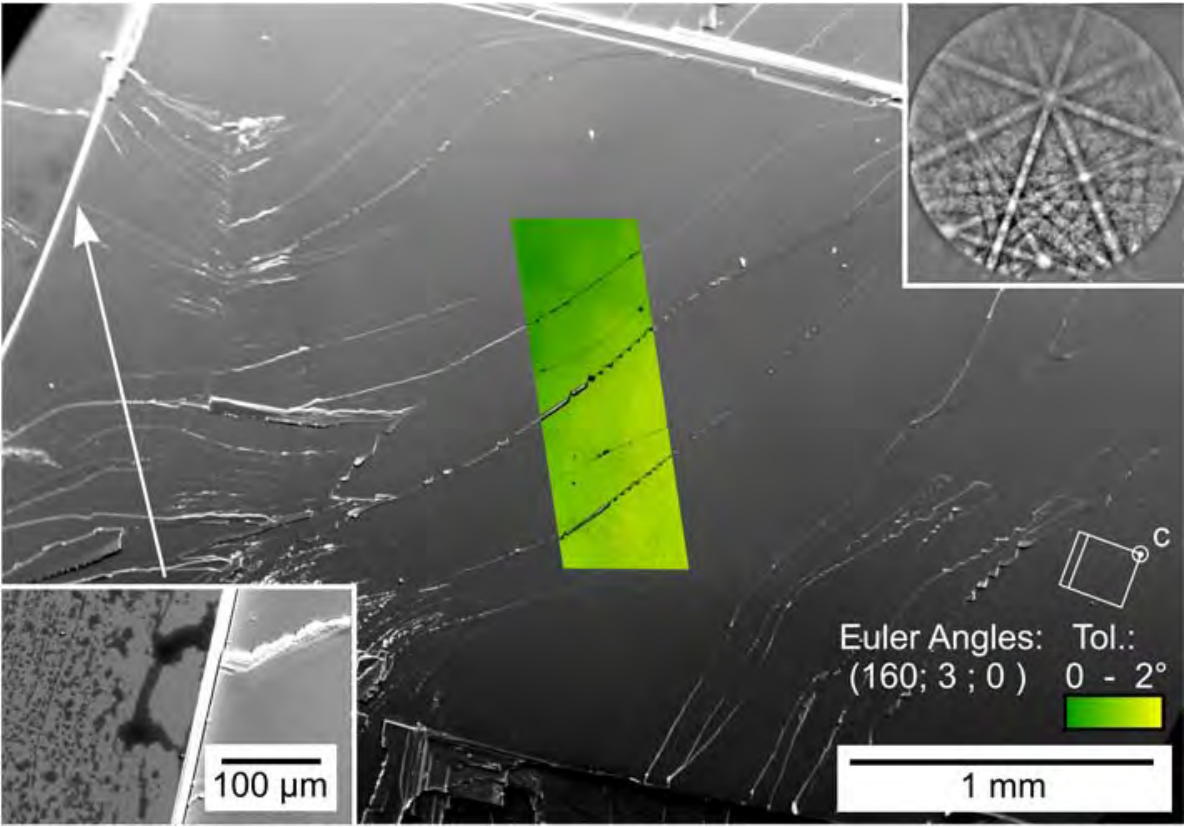}
\caption{SEM-micrograph of a crystal surface superimposed by the
orientation+IQ map of a performed EBSD-scan. The presented EBSD-pattern is
representative for patterns obtained from the extremities and centre of the
sample. The inset features the edge of the main plateau to adjacient areas
where a second phase (dark) is frequently observed.}
\label{fig:EBSD}
\end{figure}
The crystal orientation of some undoped crystals was analysed using electron backscatter diffraction (EBSD) in a Jeol JSM-7001F equipped with an EDAX Trident analysing system containing a TSL Digiview 1913 EBSD-camera. EBSD-scans were captured and evaluated using the programs TSL OIM Data Collection 5.31 and TSL OIM Analysis 5. The scans were performed using a current of about 2.40\,nA (measured with a Faraday cup) and a voltage of 20\,kV.
Figure \ref{fig:EBSD} presents a scanning electron microscopy (SEM) image of the central plateau of one sample tilted by 20\,$^{\circ}$ to enhance the topographical contrast. The EBSD-pattern presented in the figure is representative for the BaFe$_2$As$_2$ –patterns obtained from the entire extent of the sample with only minimal deviations. A material file for indexing the patterns was built using the data of ICSD-File no. 169555. The combined orientation+image quality (IQ) map of an EBSD-scan performed on the surface is superimposed on the SEM-micrograph. It shows that the crystal orientations deviate less than 2\,$^{\circ}$ from the defined Euler Angle triplet (160;3;0) illustrated by the wire frame of the unit cell (right). The crystallographic c-axis of this tetragonal crystal is hence tilted by about 3\,$^{\circ}$ from the normal of the SEM-stage.
The inset to the left shows the edge of the plateau in greater detail where a second phase enriched in carbon is frequently observed. EBSD-patterns could not be obtained from this phase indicating it may be amorphous. The inclusions of this phase were not observed on the main plateau of the crystal and are hence neglected during further analysis. The fabrication of the doped crystals is nearly identical to the undoped ones. First EBSD measurements on doped single crystals show no significant difference in crystallinity to the undoped ones. Considering that we intend to use barrier thicknesses down to 2\,nm we have to polish the surface of the crystal.
The first step was to embed the crystal in an epoxy resin and mount it on an oxidised silicon wafer to enhance handling during the three step polishing process. The surface was initially planarised by lapping in the Logitech PM5 Lapping \& Polishing System using alumina powder of 3\,$\mu$m grain size suspended in water-free isopropyl alcohol due to the sensitivity of the crystal to corrosion by water. The resulting surface showed a roughness of about 1\,$\mu$m root mean square (RMS). Subsequently, the samples were polished using decreasing grain sizes of SiC polishing foil to reduce the roughness to ca. 10\,nm RMS. The final step again performed on the PM5 system using alumina powder of 50\,nm grain size suspended in isopropyl alcohol. A relatively high load of 1.8\,kg enabled to achieve a roughness of 0.6\,nm to 2.0\,nm RMS, see \fref{fig:pol}.

\begin{figure}[htb]
\centering
\includegraphics[width=1\linewidth]{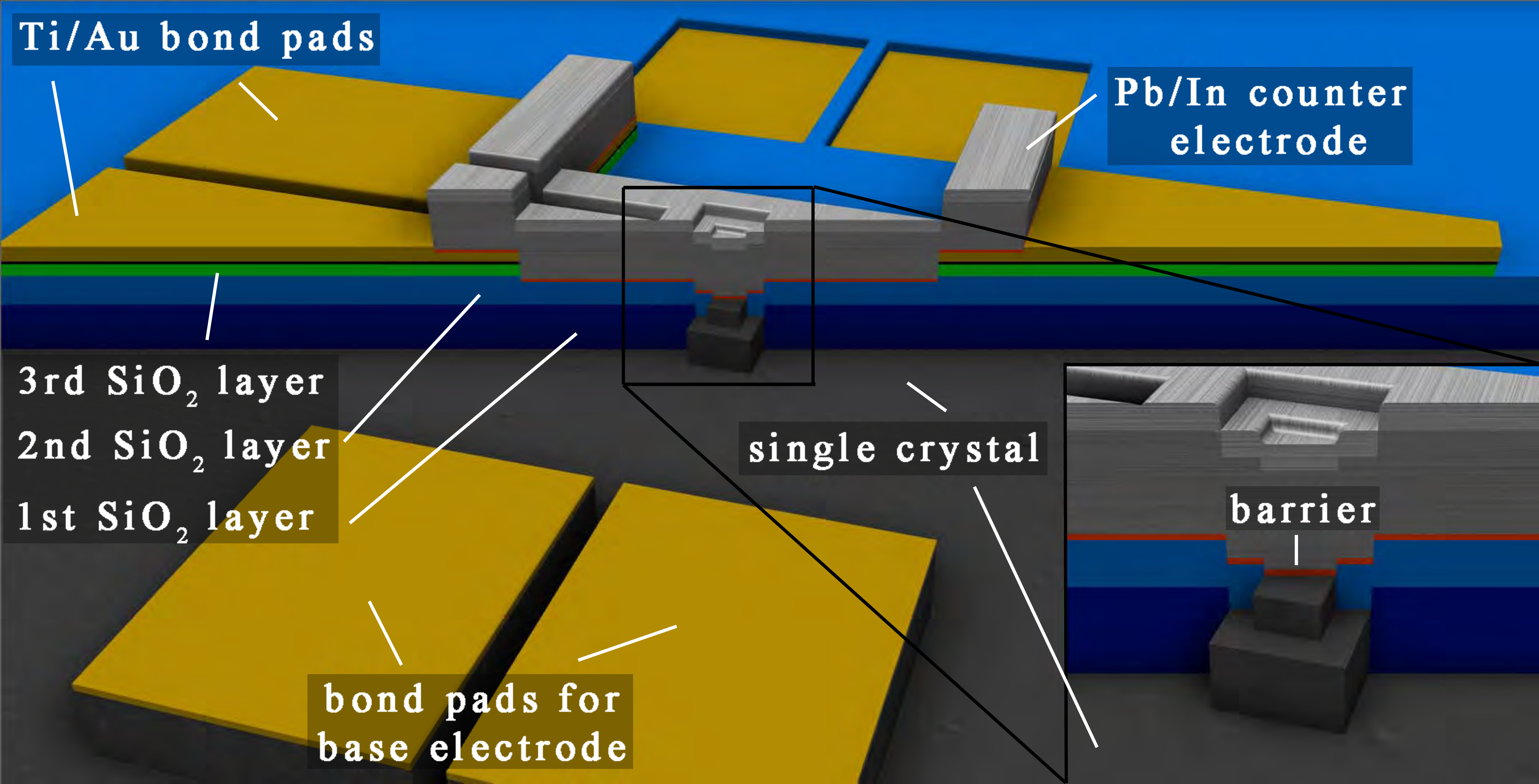}
\caption{Scheme of the junction design. The inset shows the junction interface in greater detail.}
\label{fig:schnitt}
\end{figure}

After a step of IBE to clean the surface, an Au protection layer of 50\,nm was deposited on the samples via DC sputtering directly after polishing, similar to the process used for thin film junctions \cite{Doering2012}. This avoids future degradation of the freshly polished surface during further preparation. Since the crystal interacts with normal atmosphere between cleaning and the deposition of the protection layer, we can not eliminate the possibility of surface degradation. Hence the exposure time was kept as short as possible. Additionally, the crystal surface in the junction area is cleaned again during further processing and covered with Ti in-situ. The applied junction design allows the electrical characterisation of the junction and of both electrodes in 4-probe geometry and is outlined in \fref{fig:schnitt}.

\begin{figure}[htb]
\centering
\includegraphics[width=1\linewidth]{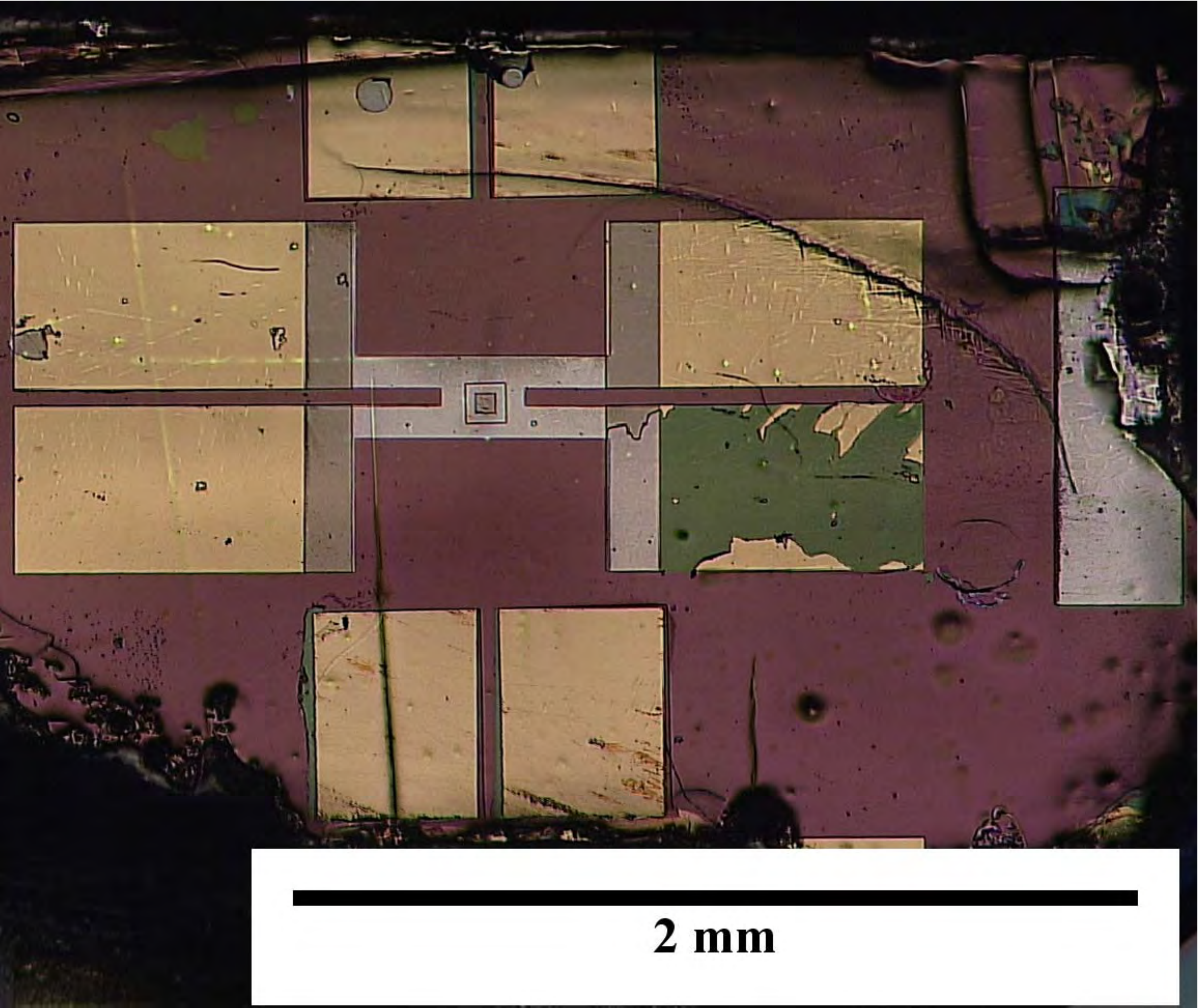}
\caption{Optical micrograph of a final junction. The distorted bond pad on the right was probably caused by surface contamination. This lead to a decreased adherence of the Ti/Au system. }
\label{fig:mic}
\end{figure}
The Ba-122 crystal acts as the base electrode, which is fully insulated from the Pb counter electrode by two layers of SiO$_2$. The Ti/Au bond pads necessary for the electrical contacting through thermosonic wirebonding are also illustrated.\\
This junction setup was achieved by first etching 100\,nm into the crystal, using an Ar ion beam (beam voltage\,=\,500\,V, fluence\,=\,0.9\,mA/cm$^{2}$) and a photo lithography mask. The Au protection layer is simultaneously removed from these regions and hence can not act as a parasitic parallel shunt on the crystal surface. A 200\,nm thick SiO$_2$ layer was subsequently deposited via reactive high frequency (HF) sputtering using a deposition rate of 2.5\,nm/min. This first insulating layer opens a big primary window for the future junction area. After a lift-off the next photo lithography mask was applied. The sample was again patterned through IBE, followed by the deposition of a second 200\,nm thick SiO$_2$ layer with a smaller window, which defines the actual junction area. After another lift-off, the mask for the bonding pads of the counter electrode was applied. Now a third SiO$_2$ layer of 100\,nm was deposited to improve the insulation between the base electrode and the bonding pads of the counter electrode. An additional Ti/Au bi-layer was deposited by DC sputtering to increase the adhesive strength of the bonding pads. The Ti layer was deposited with a deposition rate of about 0.4\,nm/s while the Au layer was deposited in-situ with a deposition rate of about 1\,nm/s. The junction interface was cleaned by IBE after the lift-off and fabrication of the counter electrode mask.\\
The existing gold layer is removed during the cleaning step and the etching time was calculated to stop at the gold/pnictide interface. Similar to the approach on high temperature superconductors \cite{Schneidewind1995} the parameters of the ion beam were adjusted to minimize the damage to the crystal surface. Nevertheless, we can not rule out a certain degree of surface modification. However, the influence of IBE processing would not be as problematic for mm-scale crystals used here as it is in thin film junctions. In any case the processing should only cause an increased effective barrier thickness. A thin Ti layer of 1.5 to 10\,nm was subsequently deposited by DC sputtering and oxidised in a normal atmosphere at a temperature of 80\,$^{\circ}$C to form a TiO$_x$ barrier. The processing of this barrier is similar to the optimised barrier described in \cite{Doering2014}.

 Finally, thermal evaporation was used to deposit the 300\,nm thick Pb counter electrode ex-situ (deposition rate\,=\,0.5\,nm/s) and an In cap layer (50\,nm) in-situ to protect the Pb layer from degradation. After a final lift-off, the complete junction as shown in \fref{fig:mic} is ready for electrical measurements.

\section{Electrical measurements}
\begin{figure}[htb]
\centering
\includegraphics[width=1\linewidth]{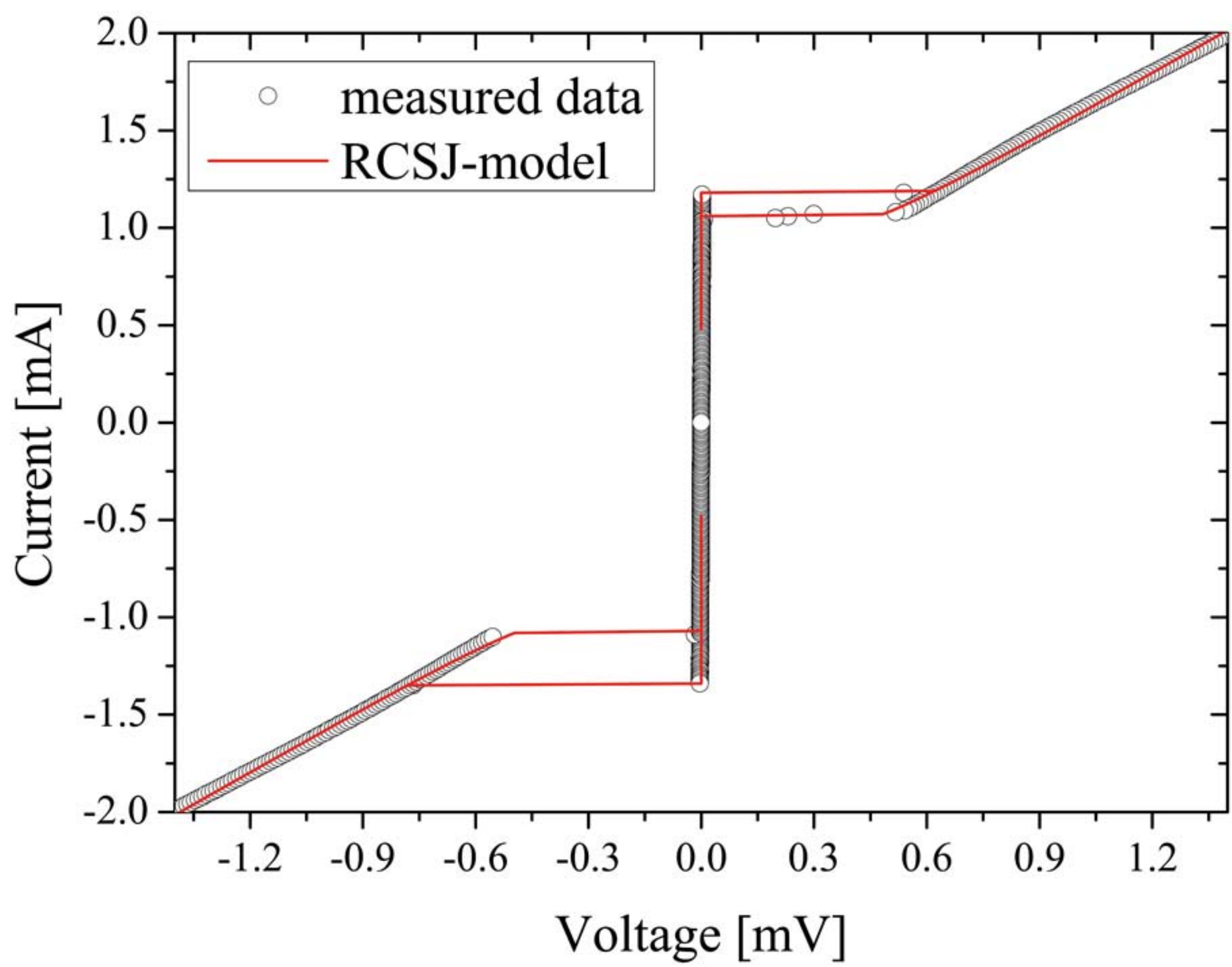}
\caption{Measured $I$-$V$ characteristic of a Josephson junction at $T=$4.2\,K compared to a fit within the RCSJ-modell. The used fitting parameters are mentioned in the text. The junction area was 15$\times$15\,$\mathrm{\mu}$m$^2$, while the thickness of the titanium oxide barrier was about 10\,nm.}
\label{fig:twox925G1-Fit}
\end{figure}

The measurements were realised in a helium dewar in which the sample was cooled to the temperature of liquid helium. An $I$-$V$ characteristic of a hysteretical Josephson junction measured at $T=$4.2\,K is presented in \fref{fig:twox925G1-Fit}.

In \fref{fig:twox925G1-Fit} we compared our measurement to the prediction of the RCSJ model \cite{McCumber1968,Stewart1968}. In order to match our data we expanded the model by an excess current. Additionally the fit was executed separately for positive and negative currents, because the measurement shows a significant asymmetry. The fit yields an $I_{\mathrm{c}}$ of 770$\mathrm{\mu}$A for positive currents and 947$\mathrm{\mu}$A for negative currents, with a symmetric excess current of 400\,$\mathrm{\mu}$A and normal state resistance $R_{\mathrm{n}}$ of 0.87\,$\Omega$. This leads to $I_{\mathrm{c}}R_{\mathrm{n}}$-products of 668\,$\mathrm{\mu}$V (positive) and 829\,$\mathrm{\mu}$V (negative). The McCumber parameters $\beta_\mathrm{C}$ are 1.76 (positive) and 3.11 (negative), respectively. Former publications on Josephson junctions with a conventional superconductor electrode \cite{Zhang2009a,Zhou2008,Schmidt2010,Doering2012b}, but also all-pnictide junctions \cite{Zhang2009,Katase2010,Katase2011,Schmidt2013}, did not show such a high $I_{\mathrm{c}}R_{\mathrm{n}}$-product. The normal resistance of this junctions is high compared to junctions on thin films with Au barriers \cite{Doering2012a,Schmidt2010} but similar to thin film junction with TiO$_x$ barriers \cite{Doering2014}. We assume that the high $I_{\mathrm{c}}R_{\mathrm{n}}$-product is caused by the influence of this insulating barrier material.
In the past we have also observed junctions with an asymmetric $I$-$V$ characteristic \cite{Doering2012b,Schmidt2010}. This asymmetry could be explained by flux trapped in the junction. Direction dependent tunnelling mechanisms including surface states \cite{Grajcar1997} and traps in the insulator \cite{Knauer1977} may also be related to an asymmetric current transport. An alternative explanation could be the presence of a time-reversal symmetry broken state in pnictide superconductors as described in \cite{Huang2014}.
\begin{figure}[htb]
\centering
\includegraphics[width=1\linewidth]{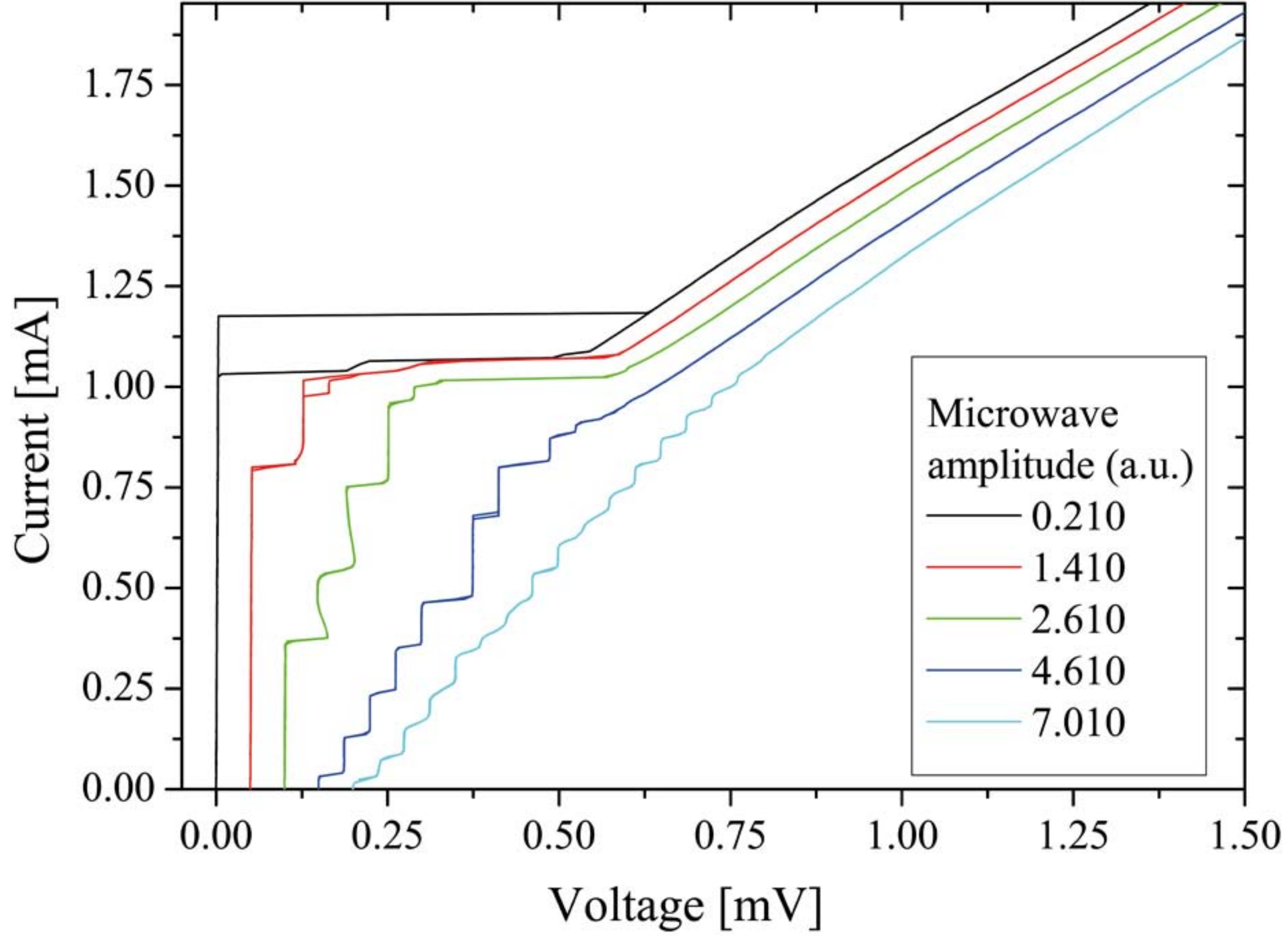}
\caption{Positive branch of the I-V-characteristic shown in \fref{fig:twox925G1-Fit} under microwave irradiation ($f$=18\,GHz). The curves are shifted by 50\,$\mu$V against each other.}
\label{fig:twox925G1-micro}
\end{figure}
In \fref{fig:twox925G1-micro} we show the influence of microwave irradiation on the junction. A decrease of the critical Josephson current as well as the formation of Shapiro steps are clearly visible at voltage values of 37\,$\mathrm{\mu}$V and integer multiples, corresponding to the irradiated frequency of 18\,GHz. Thus, an actual presence of a Josephson effect in the junction can be assumed.
Some of the Steps depicted in \fref{fig:twox925G1-micro} show an unusual behavior in which the step is tilted. This is probably due to the low irradiation frequency, resulting in an $\Omega$ of 0.044 for negative and 0.055 for positive bias were $\Omega$ is the ratio of the frequency of irradiation and the characteristic frequency of the contact. A similar behavior was observed in \cite{Kuznik1988}.

\section{Summary}
We described a polishing process that enables us to achieve a high quality surface on Ba-122 single crystals. On this surface we can prepare Josephson junctions using the classical thin film technologies photo lithography, IBE and sputtering. The junction layout we used was derived from earlier thin film junctions and will be optimised for single crystals in the future. Furthermore we presented first electrical measurements of the Josephson effect on these junctions, where a high $I_{\mathrm{c}}R_{\mathrm{n}}$-product could be observed.
Detailed investigations of junction properties and their dependence on barrier parameters are the subject of current research.

\section{Acknowledgements}
This work was funded within the European Community project IRON-SEA under grant number FP7-283141 and by the German Research Community (DFG) within priority program 1458 under grant number SE664/15-2. S.~Schmidt was funded by the Landesgraduiertenf\"orderung Th\"uringen and N.~Hasan was funded by the German Academic Exchange Service (DAAD).

\section*{References}
\bibliographystyle{iopart-num}
\bibliography{Literatur}

\end{document}